%% file: chaty_mykonos.tex
\documentclass[
    ,final            
  ]
  {aipproc}

\layoutstyle{8x11single}

\input{symbols}

\begin{document}

\title{Multi-wavelength study of High Mass X-ray Binaries}


\author{S. Chaty}{
  address={Laboratoire AIM (UMR 7158 CEA/DSM-CNRS-Universit\'e Paris Diderot),
Irfu/Service d'Astrophysique, CEA-Saclay,
FR-91191 Gif-sur-Yvette Cedex, France}
}

\begin{abstract}
The INTEGRAL satellite has revealed a major population of supergiant
High Mass X-ray Binaries in our Galaxy, revolutionizing our
understanding of binary systems and their evolution. This population,
constituted of a compact object orbiting around a massive and luminous 
supergiant star, exhibits unusual properties, either being extremely absorbed, 
or showing very short and intense flares.  An intensive set of
multi-wavelength observations has led us to reveal their nature, and
to show that these systems are wind-fed accretors, closely related to
massive star-forming regions.  In this paper I describe the
characteristics of these sources, showing that this newly
revealed population is closely linked to the evolution of active and massive 
OB stars with a compact companion. The last section emphasizes the formation and evolution of such High Mass X-ray Binaries hosting a supergiant star.
\end{abstract}

\maketitle


\section{The $\gamma$-ray sky seen by the {\it INTEGRAL} satellite}

The {\it INTEGRAL} observatory is an ESA satellite launched on 17 October
2002 by a PROTON rocket on an excentric orbit. It is hosting 4
instruments: 2 $\gamma$-ray coded-mask telescopes --the imager IBIS
and the spectro-imager SPI, observing in the range 10 keV-10 MeV, with
a resolution of $12\amin$ and a field-of-view of $19\adeg$-- a
coded-mask telescope JEM-X (3-100 keV), and an optical telescope
(OMC).


The $\gamma$-ray sky seen by {\it INTEGRAL} is very rich, since 723 sources
have been detected by {\it INTEGRAL}, reported in the $4^{th}$ IBIS/ISGRI soft
$\gamma$-ray catalogue, spanning nearly 7 years of observations in the 17-100
keV domain \cite{bird:2010}.  
%
Among these sources, there are 185 X-ray binaries (representing 26\% of the whole sample of sources detected by {\it INTEGRAL}, called ``IGRs'' in the following), 255 Active Galactic Nuclei (35\%), 35 Cataclysmic Variables (5\%), and $\sim 30$ sources of other type (4\%): 15 SuperNova Remnants, 4 Globular Clusters, 3 Soft $\gamma$-ray Repeaters, 2 $\gamma$-ray Burst, etc. 
215 objects still remain unidentified (30\%).  X-ray binaries are separated in 95 Low Mass X-ray Binaries (LMXBs) and 90 High Mass X-ray Binaries (HMXBs), each category representing $\sim 13$\% of IGRs.
Among identified HMXBs, there are 24 BeHMXBs (HMXBs hosting a Be companion star) and 19 sgHMXBs (HMXBs hosting a supergiant companion star), representing respectively 31\% and 24\% of HMXBs).

It is interesting to follow the evolution of the ratio between BeHMXBs
and sgHMXBs.  During the pre-{\it INTEGRAL} era, HMXBs were mostly BeHMXB
systems.  For instance, in the catalogue of 130 HMXBs by
\citet{liu:2000}, there were 54 BeHMXBs and 7 sgHMXBs (respectively 42\% and 5\% of the total number of
HMXBs).  Then, the situation changed drastically with the first
HMXBs identified by {\it INTEGRAL}: in the catalogue of 114 HMXBs (+128 in the Magellanic Clouds) of \citet{liu:2006}, there
were 60\% of BeHMXBs and 32\% of sgHMXBs firmly identified.  Therefore,
while the ratio of BeHMXBs/HMXBs increased by a factor of 1.5 only, the
sgHMXBs/HMXBs ratio increased by a factor of 6.


\subsection{Let the  {\it INTEGRAL} show go on!}

The ISGRI detector on the IBIS imager has performed a detailed survey of the
Galactic plane, discovering
many new high energy celestial objects, most of which 
reported in \citet{bird:2010}\footnote{See an up-to-date list at 
{\em http://irfu.cea.fr/Sap/IGR-Sources/}, maintained by J. Rodriguez and A. Bodaghee}.  The
most important result of {\it INTEGRAL} to date is the discovery of
many new high energy sources -- concentrated in the Galactic plane,
mainly towards tangential directions of Galactic arms, rich in star forming
regions, -- exhibiting common characteristics which previously had
rarely been seen (see e.g. \citeauthor{chaty:2005a}
\citeyear{chaty:2005a}). 
Many of them are HMXBs hosting a neutron star (NS) orbiting around an OB companion, in most cases a supergiant star. 
Nearly all the {\it INTEGRAL} HMXBs for which both spin and orbital
periods have been measured are located in the upper part of the Corbet
diagramme \citep{corbet:1986} (see Figure \ref{liufig}).
They are wind accretors, typical of
sgHMXBs, and X-ray pulsars exhibiting longer pulsation
periods and higher absorption (by a factor $\sim4$) as compared to the
average of previously known HMXBs \citep{bodaghee:2007}. 
They divide into two classes: some are very obscured, exhibiting a huge intrinsic and
local extinction, --the most extreme example being the highly absorbed
source IGR~J16318-4848 \citep{filliatre:2004}--, and the others are
HMXBs hosting a supergiant star and exhibiting fast and transient
outbursts -- an unusual characteristic among HMXBs.  These are
therefore called Supergiant Fast X-ray Transients (SFXTs,
\citeauthor{negueruela:2006a} \citeyear{negueruela:2006a}),
with IGR~J17544-2619
being their archetype \citep{pellizza:2006}.

\subsection{Multi-wavelength observations of  {\it INTEGRAL} sources} \label{observations-IGRs}

To better characterise this population, \citet{chaty:2008} and \citet{rahoui:2008} 
studied a sample of 21 IGRs
belonging to both classes described above.  Sources of
this sample are X-ray pulsars, with high $P_\mathrm{spin}$ from 139 to 5880\,s
and $\Porb$ ranging from 4 to 14\,days.  They are therefore wind
accreting sgHMXBs, according to the Corbet diagramme 
(Figure \ref{liufig}). Multiwavelength observations were performed from 2004 to 2008 at the
European Southern Observatory (ESO), using Target of Opportunity (ToO)
and Visitor modes, in 3 domains: optical ($400-800$\,nm) with EMMI,
NIR ($1-2.5 \microns$) with SOFI, both instruments at the focus of the
3.5m New Technology Telescope (NTT) at La Silla, and mid-infrared
(MIR, $5-20 \microns$) with the VISIR instrument on Melipal, the 8m
Unit Telescope 3 (UT3) of the Very Large Telescope (VLT) at Paranal
(Chile). They also used data from the GLIMPSE survey of {\it Spitzer}.  With
these observations they performed accurate astrometry, identification,
photometry and spectroscopy,
aiming at identifying IGR counterparts and the nature of the
companion star, deriving their distance, and finally characterising
the presence and temperature of their circumstellar medium, by fitting
their spectral energy distribution (SED).  

The main results of this study are that 15 of
these IGRs are identified as HMXBs, and among them 12 HMXBs contain
massive and luminous early-type companion stars. By combining optical,
NIR and MIR photometry, and fitting their SEDs, \citet{rahoui:2008}
showed that (i) most of these sources exhibit an intrinsic absorption
and (ii) three of them exhibit a MIR excess, which they suggest to be
due to the presence of a cocoon of dust and/or cold gas enshrouding
the whole binary system, with a temperature of $T_d \sim 1000$\,K,
extending on a radius of $R_d \sim 10\Rstar$ (see \citeauthor{chaty:2006c} \citeyear{chaty:2006c}).

\section{Supergiant Fast X-ray Transients}

\subsection{General characteristics}

SFXTs constitute a new class of $\sim 12$ sources identified among the
recently discovered IGRs. They are HMXBs hosting NS orbiting around sgOB companion stars, exhibiting peculiar characteristics compared to ``classical'' HMXBs: rapid outbursts lasting only for hours, faint quiescent emission, and high energy spectra requiring a black hole (BH) or NS accretor. The flares rise in tens of minutes, last for $\sim$ 1 hour, their frequency is $\sim7$\,days, and their luminosity $L_x$ reaches $\sim 10^{36} \ergs$ at the outburst peak.

\subsection{IGR~J17544-2619, archetype of SFXTs}

\begin{figure}
  \includegraphics[height=.38\textheight,angle=-90]{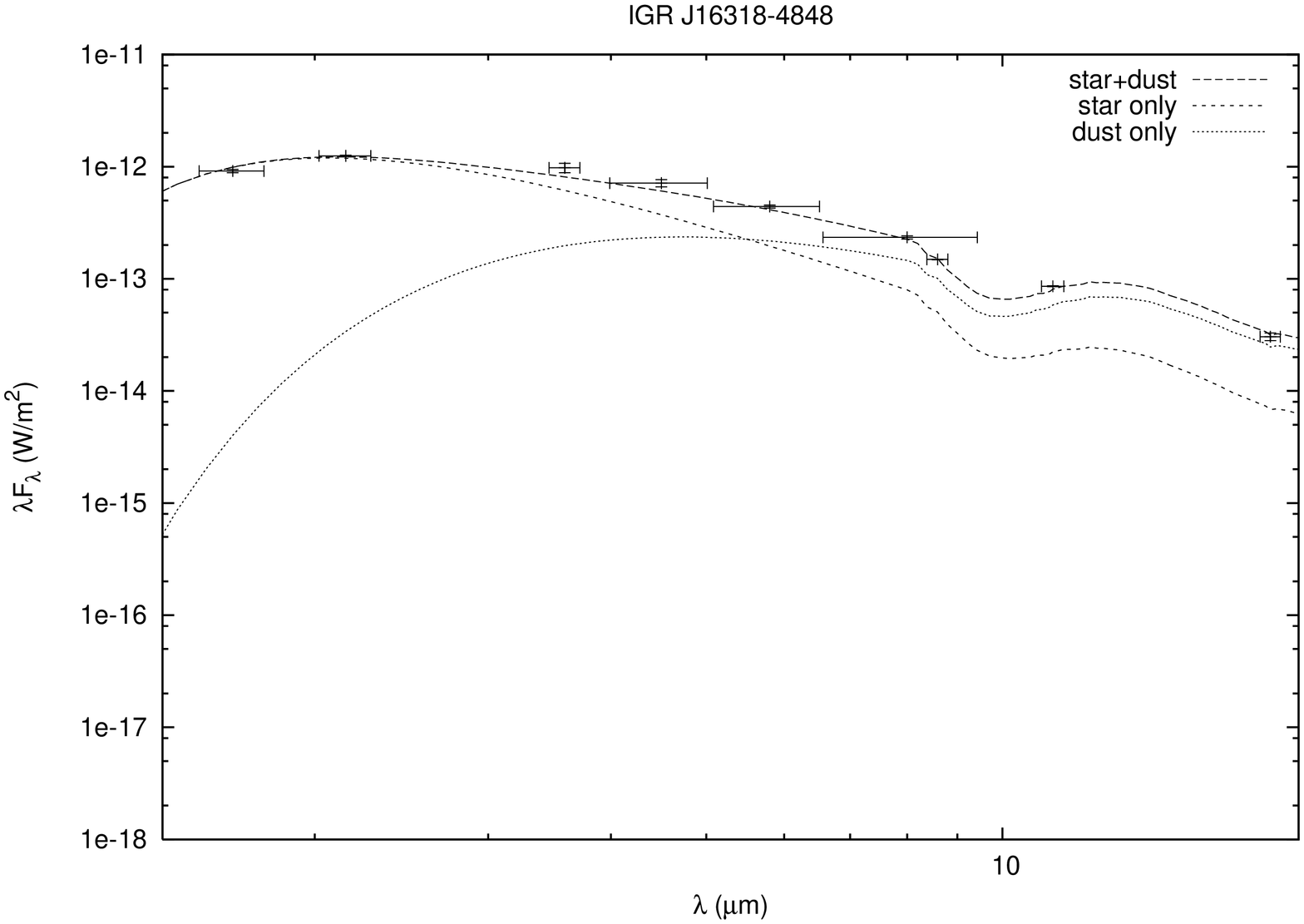}
  \includegraphics[height=.395\textheight,angle=-90]{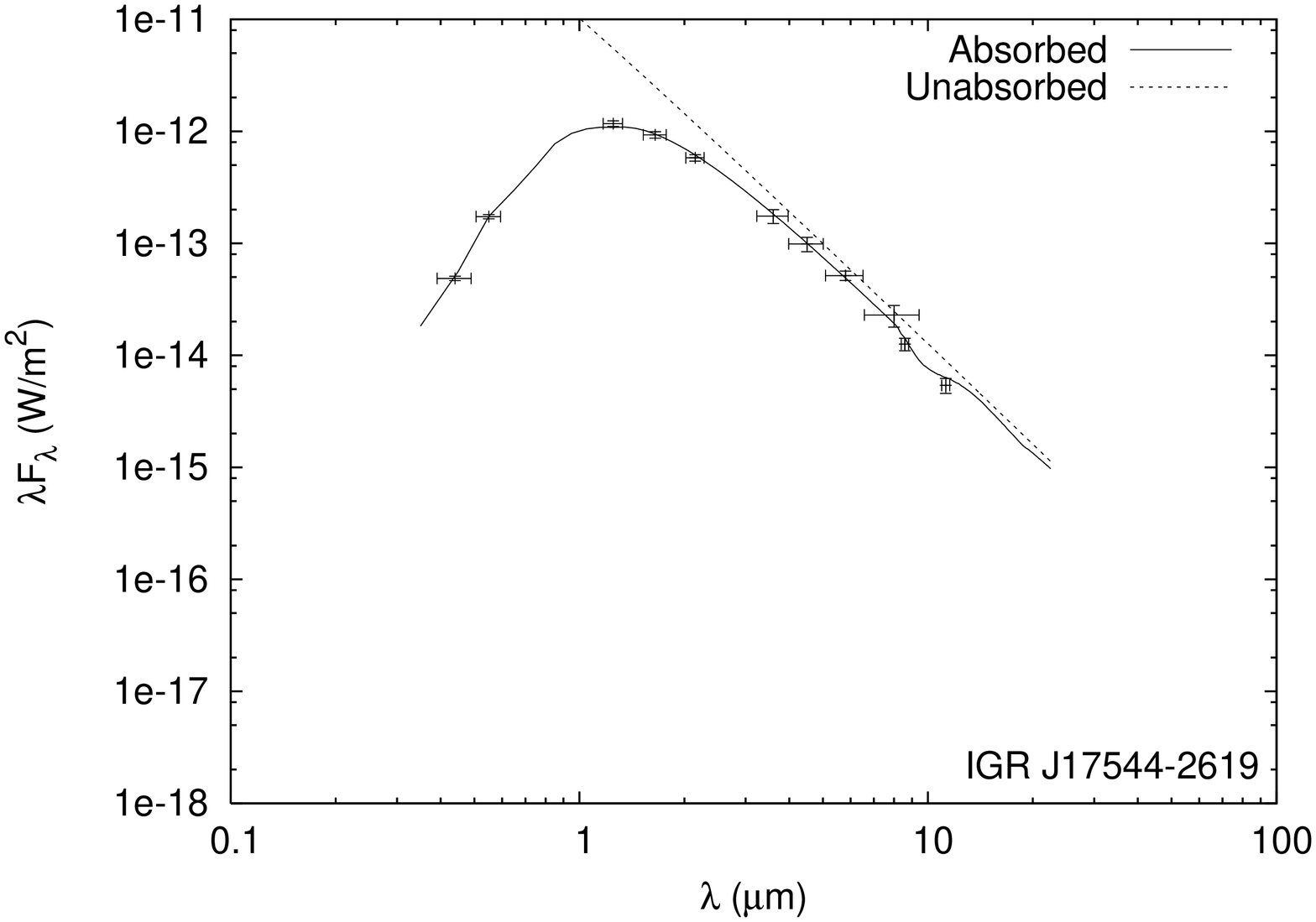}
  \caption{\label{figure:igrj16318-igrj17544} Optical to MIR SEDs of
  IGR~J16318-4848 (left) and IGR~J17544-2619 (right), including data
  from ESO/NTT, VISIR on VLT/UT3 and {\it Spitzer} \citep{rahoui:2008}.
  IGR~J16318-4848 exhibits a MIR excess, interpreted as the signature of a strong stellar outflow
  coming from the sgB[e] companion star \citep{filliatre:2004}.  On the
  other hand, IGR~J17544-2619 is well fitted with only a stellar
  component corresponding to the O9Ib companion star spectral type
  \citep{pellizza:2006}.}
\end{figure}

   This bright recurrent transient X-ray source was discovered by {\it
     INTEGRAL} on 17 September 2003 \citep{sunyaev:2003b}. {\it
     XMM-Newton} observations showed that it exhibits a very hard
   X-ray spectrum, and a relatively low intrinsic absorption ($\nh
   \sim 2 \times 10^{22}\cmmoinsdeux$,
   \citeauthor{gonzalez-riestra:2004}
   \citeyear{gonzalez-riestra:2004}).  Its bursts last for hours, and
   inbetween bursts it exhibits long quiescent periods, which can
   reach more than 70\,days. The X-ray behaviour is complex on long,
   mean and short-term timescales: rapid flares are detected by {\it
     INTEGRAL} on all these timescales, on pointed and 200s binned
   lightcurve (Zurita Heras \& Chaty in prep.). The compact object is
   probably a NS \citep{intzand:2005}.  \citet{pellizza:2006} managed
   to get optical/NIR ToO observations only one day after the
   discovery of this source. They identified a likely counterpart
   inside the {\it XMM-Newton} error circle, confirmed by an accurate
   localization from {\it Chandra}.  Spectroscopy showed that the
   companion star was a blue supergiant of spectral type O9Ib, with a
   mass of $25-28 \Msol$, a temperature of $T\sim 31000$~K, and a
   stellar wind velocity of $265 \pm 20 \kms$ (which is faint for O
   stars): the system is therefore an HMXB \citep{pellizza:2006}.
   \citet{rahoui:2008} combined optical, NIR and MIR observations and
   showed that they could accurately fit the observations with a model
   of an O9Ib star, with a temperature $\Tstar \sim 31000$~K and a
   radius $\Rstar = 21.9 \Rsol$. They derived an absorption A$_v = 6.1
   \mags$ and a distance D~$=3.6$~kpc. Therefore the source does not
   exhibit any MIR excess, and is well fitted by an unique stellar
   component (see Figure \ref{figure:igrj16318-igrj17544}, right
   panel, \citeauthor{rahoui:2008} \citeyear{rahoui:2008}).

\subsection{Classification of SFXTs}

We can divide the SFXTs in two groups, according to the duration and
frequency of their outbursts, and their
$\frac{L_\mathrm{max}}{L_\mathrm{min}}$ ratio.  The classical SFXTs
exhibit a very low quiescence $L_X$ and a high variability, while
intermediate SFXTs exhibit a higher $<L_X>$, a lower
$\frac{L_\mathrm{max}}{L_\mathrm{min}}$ and a smaller variability,
with longer flares.  SFXTs might appear like persistent sgHMXBs with
$<L_X>$ below the canonical value of $\sim 10^{36} \ergs$, and flares
superimposed.  But there might be some observational bias in these
general characteristics, therefore the distinction between SFXTs and
sgHMXBs is not well defined yet.
While the typical hard X-ray variability factor (the ratio between
deep quiescence and outburst flux) is less than 20 in
classical/absorbed systems, it is higher than 100 in SFXTs (some
sources can exhibit flares in a few minutes, like for instance
XTE\,J1739-302 \& IGR\,J17544-2619).  The
intermediate SFXTs exhibit smaller variability factors. \\

\subsubsection{SFXT behaviour: clumpy wind accretion?}

Such sharp rises exhibited by SFXTs are incompatible with the orbital
motion of a compact object through a smooth medium
(\citeauthor{negueruela:2006a} \citeyear{negueruela:2006a},
\citeauthor{smith:2006} \citeyear{smith:2006},
\citeauthor{sguera:2005} \citeyear{sguera:2005}).  Instead, flares
must be created by the interaction of the accreting compact object
with the dense clumpy stellar wind (representing a large fraction of
stellar $\Mdot$).  In this
case, the flare frequency depends on the system geometry, and the
quiescent emission is due to accretion onto the compact object of
diluted inter-clump medium, explaining the very low quiescence level
in classical SFXTs. \\

\subsubsection{Macro-clumping scenario}

Each SFXT outburst is due to the accretion of a single clump, assuming that 
the X-ray lightcurve is a direct tracer of the wind density distribution.
The typical parameters in this scenario are:
a compact object with large orbital radius: $10 \Rstar$,
a clump size of a few tenths of $\Rstar$,
a clump mass of $10^{22-23}g$ (for $\nh=10^{22-23}\cmmoinsdeux$),
a mass loss rate of $10^{-(5-6)} \Msol/yr$,
a clump separation of order $R_{\star}$ at the orbital radius,
and a volume filling factor: 0.02$->$0.1.
The flare to quiescent count rate ratio is directly related to the $\frac{clump}{inter-clump}$ density ratio, which ranges between
15-50 for intermediate SFXTs, and $10^{2-4}$ for "classical" SFXTs.
A very high degree of porosity (macroclumping) is required to reproduce
the observed outburst frequency in SFXTs, in good agreement with UV line
profiles and line-driven instabilities at large radii 
(\citeauthor{oskinova:2007} \citeyear{oskinova:2007};
\citeauthor{runacres:2005} \citeyear{runacres:2005}; 
\citeauthor{walter:2007} \citeyear{walter:2007}). \\

\subsubsection{SFXTs in the context of sgHMXBs}

To explain the emission of SFXTs in the context of sgHMXBs, 
\citet{negueruela:2008} and \citet{walter:2007} invoke the
existence of two zones around the supergiant star, of high and low
clump density respectively.  This would naturally explain the smooth
transition between sgHMXBs and SFXTs, and the existence of intermediate
systems; the main difference between classical sgHMXBs and SFXTs being
in this scenario the NS orbital radius.
Indeed, a basic model of porous wind predicts a substantial change in the
properties of the wind "seen by the NS" at a distance $r \sim 2
\Rstar$ (\citeauthor{negueruela:2008} \citeyear{negueruela:2008}), where we
stop seeing persistent X-ray sources. There are 2-regimes:
either the NS sees a large number of clumps, because it is
  embedded in a quasi-continuous wind;
or the number density of clumps is so small that the NS is effectively 
orbiting in an empty space.

The observed division between sgHMXBs (persistent sgHMXBs and SFXTs)
is therefore naturally explained by simple geometrical differences
in the orbital configurations:

\begin{enumerate}

\item The obscured sgHMXBs (persistent and luminous systems) would have 
short and circular orbits lying 
inside the zone of stellar wind high clump density ($R_{orb} \sim 2\Rstar$).

\item The intermediate SFXTs would have short orbits, circular or eccentric, 
and possible periodic outbursts, the NS being inside the narrow transition zone.

\item The classical SFXTs would have larger and eccentric orbital radius,
the NS orbiting outside the high density zone.

\end{enumerate}

\subsubsection{IGR\,J18483-0311: an intermediate SFXT?}

X-ray properties of this system were suggesting an SFXT nature
\citep{sguera:2007}, exhibiting however an unusual behaviour: its
outbursts last for a few days (to compare to hours for classical
SFXTs), and the ratio $L_{max}/L_{min}$ only reaches $\sim 10^3$ (meaning that its quiescence is at a higher level than the ratio $\sim 10^4$ for classical
SFXTs). Moreover, its orbital period $\Porb$=18.5d is low compared to
classical SFXTs (with large/eccentric orbits). Finally, its orbital
and spin periods ($P_\mathrm{spin}$=21.05s) located it ambiguously inbetween Be and
sgHMXBs in the Corbet Diagramme (see Figure \ref{liufig}).
\citet{rahoui:2008a} identified the companion star of this system
as a B0.5Ia supergiant, unambiguously showing that this system is an
SFXT.  Furthermore, they suggest that this system could be the first
firmly identified intermediate SFXT, characterised by short, eccentric
orbit (with an eccentricity $e$ between 0.4 and 0.6),
and long outbursts... An "intermediate"
SFXT nature would explain the unusual characteristics of this source among
"classical" SFXTs.

\subsubsection{What is the origin of ``misplaced'' sgHMXBs?}

\begin{figure}
\includegraphics[height=.3\textheight,angle=0]{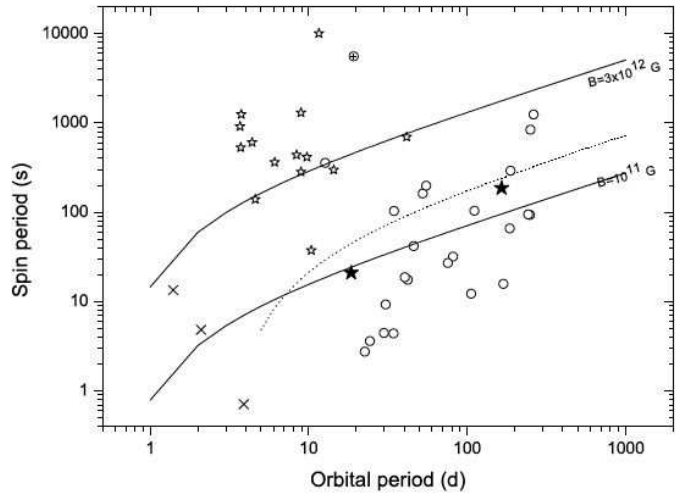}
\caption{\label{liufig}
Corbet diagram, adapted and taken from \citet{liu:2010}, showing the relation between orbital and spin periods. The open circles are for Be/X-ray binaries and the open stars for wind-fed sgHMXBs. The two ``misplaced'' SFXTs --IGR\,J11215-5952 and IGR\,J18483-0311-- are indicated by the filled stars. The solid lines stand for the theoretical equilibrium period for sgHMXBs, with magnetic field of $10^{11}$ and $3 \times 10^{12}$\,G, respectively. The dot line is the theoretical line with the parameters of B1 Ia supergiants (see \citeauthor{liu:2010} \citeyear{liu:2010} for more details).
}
\end{figure}

As noted by \citet{liu:2010}, there are two ``misplaced'' SFXTs in the Corbet diagram: IGR\,J11215-5952 (a neutron star orbiting a B1 Ia star, \citeauthor{negueruela:2005b} \citeyear{negueruela:2005b}) and IGR\,J18483-0311, described in the previous paragraph (see Figure \ref{liufig}). According to \citet{liu:2010}, these 2 SFXTs can not have evolved from normal Main Sequence O-type stars, since they are not at the equilibrium spin period of sgHMXBs (see e.g. \citeauthor{waters:1989} \citeyear{waters:1989}). They must therefore be the descendants of BeHMXBs (i.e. hosting O-type emission line stars), after the NS has reached the equilibrium spin period \citep{liu:2010}.

\section{Obscured HMXBs}

  \subsection{IGR~J16318-4848, an extreme case}

  IGR~J16318-4848 was the first source discovered by IBIS/ISGRI on
  {\it INTEGRAL} on 29 January 2003 \citep{courvoisier:2003}, with a
  $2 \amin$ uncertainty.  {\it XMM-Newton} observations revealed a
  comptonised spectrum exhibiting an unusually high level of
  absorption: $\nh \sim 1.84 \times 10^{24} \cmmoinsdeux$
  \citep{matt:2003}.  The accurate localisation by {\it XMM-Newton}
  allowed \citet{filliatre:2004} to rapidly trigger ToO photometric and
  spectroscopic observations in optical/NIR, leading to the
  confirmation of the optical counterpart \citep{walter:2003} and to
  the discovery of the NIR one \citep{filliatre:2004}.  The extremely
  bright NIR source (B\,$>25.4\pm1$; I\,$=16.05\pm0.54$, J\,$= 10.33\pm 0.14$;
  H\,$=8.33\pm 0.10$ and Ks\,$=7.20 \pm 0.05 \mags$) exhibits an
  unusually strong intrinsic absorption in the optical ($A_v = 17.4
  \mags$), 100 times stronger than the interstellar absorption along
  the line of sight ($A_v = 11.4 \mags$), but still 100 times lower
  than the absorption in X-rays.  This led \citet{filliatre:2004} to
  suggest that the material absorbing in X-rays was concentrated
  around the compact object, while the material absorbing in
  optical/NIR was enshrouding the whole system.  The NIR spectroscopy
  in the $0.95-2.5 \microns$ domain allowed them to identify the nature of
  the companion star, by revealing an unusual spectrum, with many
  strong emission lines:

\begin{enumerate}

\item H, He${\rm I}$ (P-Cyg) lines:
  characteristic of dense/ionised wind at v\,$=400$\,km/s,

\item He${\rm II}$ lines: 
  the signature of a highly excited region,

\item $[$Fe${\rm II}]$ lines:
  reminiscent of shock heated matter,

\item Fe${\rm II}$ lines:
  emanating from media of densities $>10^5-10^6$\,cm$^{-3}$,

\item Na${\rm I}$ lines:
  coming from cold/dense regions.

\end{enumerate}

All these lines originate from a highly complex, stratified
circumstellar environment of various densities and temperatures,
suggesting the presence of an envelope and strong stellar outflow
responsible for the absorption. Only luminous early-type stars such as
sgB[e] show such extreme environments, and
\citet{filliatre:2004} concluded that IGR~J16318-4848 was an unusual
HMXB hosting a sgB[e] with characteristic luminosity of
$10^6 \Lsol$ and mass of $30 \Msol$, located at a distance
between 1 and 6 kpc (see also \citeauthor{chaty:2005a} \citeyear{chaty:2005a}).
This source would therefore be the second HMXB hosting a sgB[e] star,
after CI Cam (see \citeauthor{clark:1999} \citeyear{clark:1999}). 

The question of this huge absorption was still pending, 
and only MIR observations would allow to solve this question,
and understand its origin.
By combining optical, NIR and MIR observations, and
fitting these observations with a model of sgB[e] companion star,
\citet{rahoui:2008} showed that IGR~J16318-4848 was exhibiting a MIR
excess (see Figure \ref{figure:igrj16318-igrj17544}, left panel), that they
interpreted as due to the strong stellar outflow emanating from
the sgB[e] companion star.  They found that the companion star had a
temperature of $\Tstar=22200$\,K and radius $\Rstar = 20.4 \Rsol = 0.1$\,a.u., 
consistent with a supergiant star, and
an extra component of temperature T $=1100$\,K and radius R\,$= 11.9\Rstar 
= 1$\,a.u., with A$_v = 17.6 \mags$. 
%
%
Recent MIR spectroscopic observations with VISIR at the VLT showed
that the source was exhibiting strong emission lines of H, He, Ne, PAH, Si,
proving that the extra absorbing component was made of dust and
cold gas.

By assuming a typical orbital period of 10\,days and a mass of the
companion star of $20 \Msol$, we obtain an orbital separation of $50
\Rsol$, smaller than the extension of the extra component of dust/gas
($= 240 \Rsol$),
suggesting that this dense and absorbing circumstellar material envelope enshrouds the whole binary system, like 
a cocoon (see Figure \ref{figure:obscured-sfxt}, left panel).
We point out that this source exhibits such extreme
characteristics that it might not be fully representative of the other
obscured sources.

\section{The Grand Unification: different geometries, different scenarios}

In view of the results described above, 
there seems to be a continuous trend, from classical and/or absorbed
sgHMBs, to classical SFXTs. We outline in the following this trend.

\begin{figure}
\includegraphics[height=.275\textheight,angle=0]{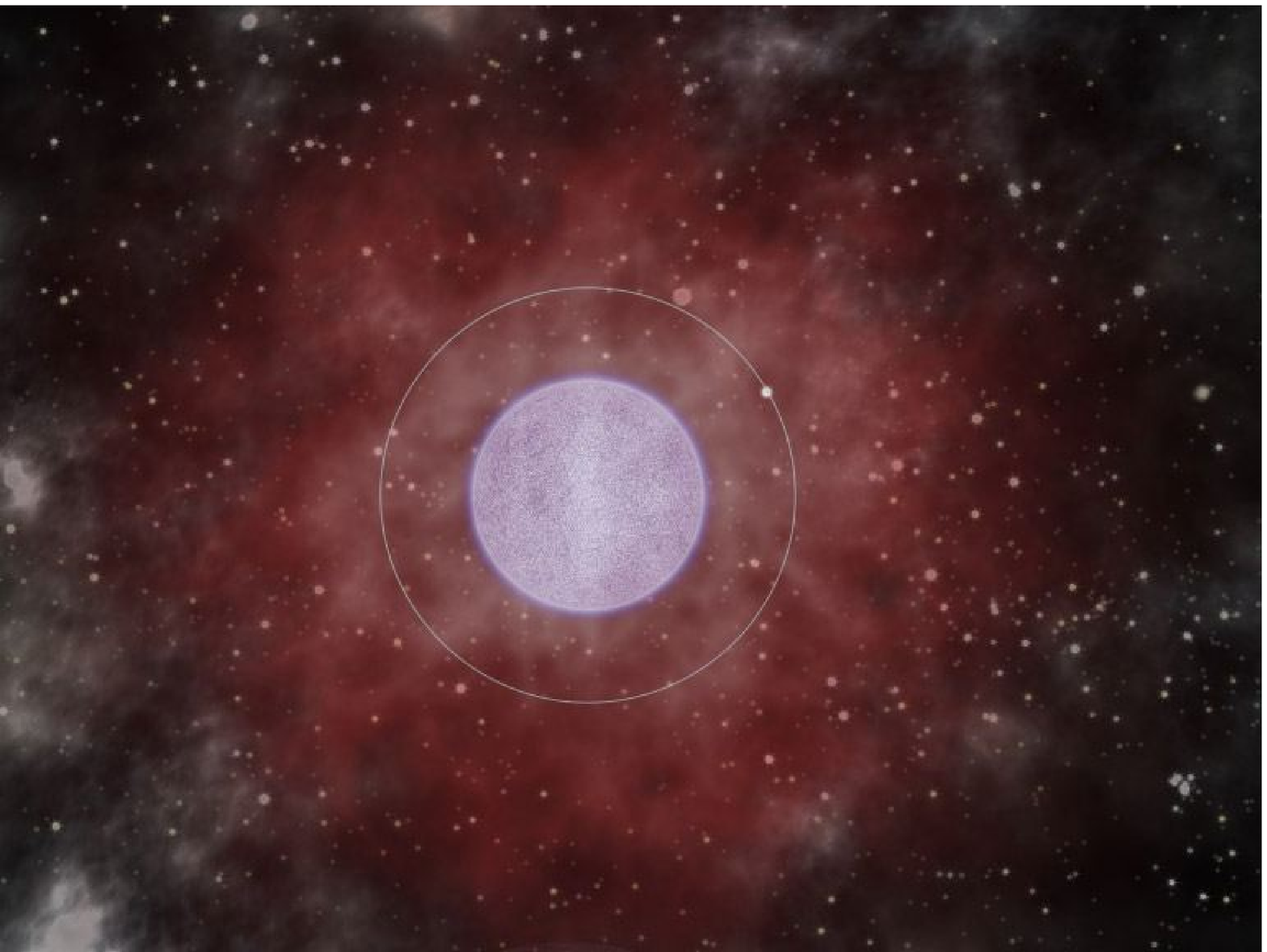}
\includegraphics[height=.275\textheight,angle=0]{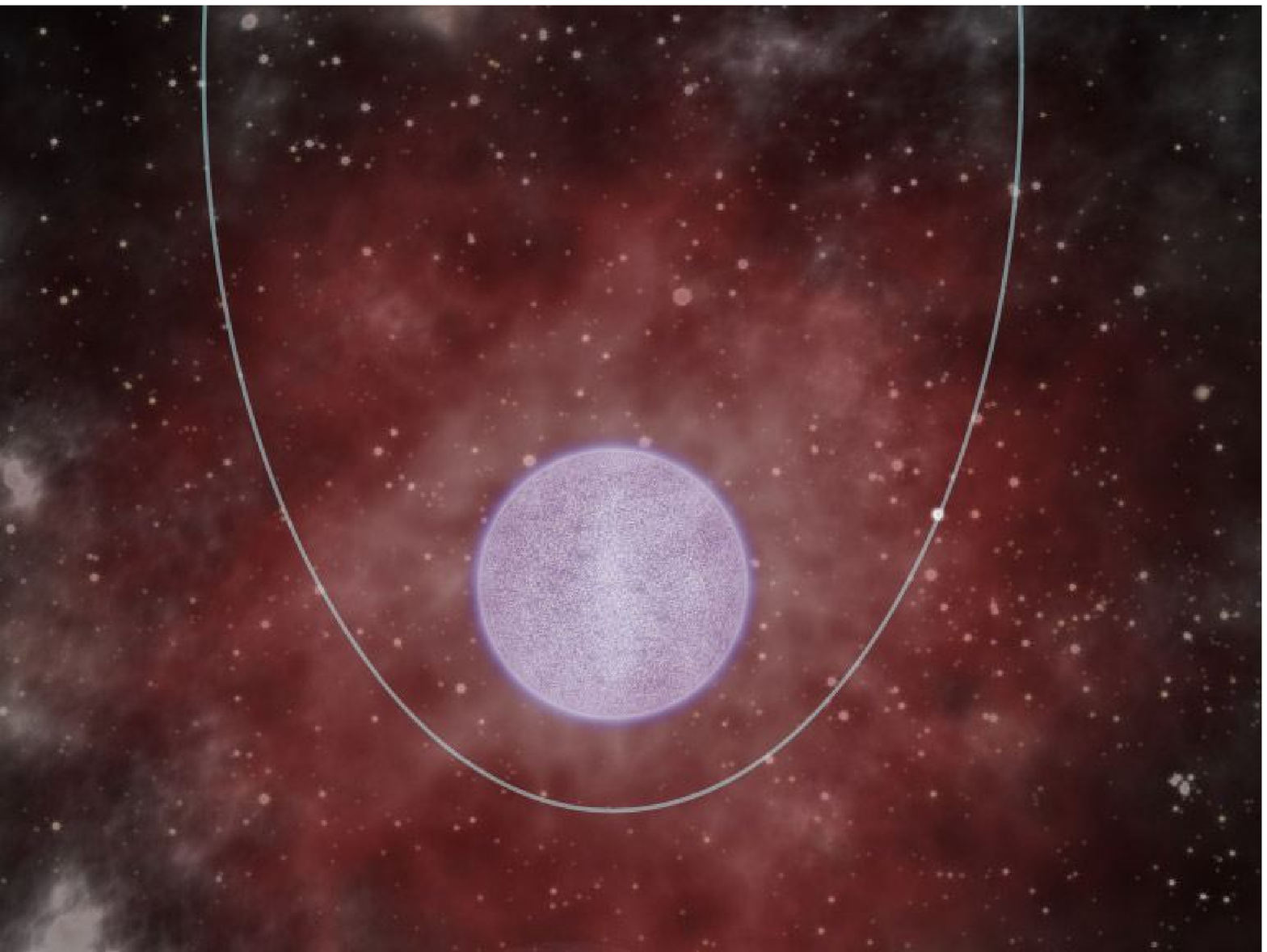}
\caption{\label{figure:obscured-sfxt}
  Scenarios illustrating two possible configurations of {\it INTEGRAL} 
  sources: a NS orbiting a supergiant
  star on a circular orbit (left image); and on an eccentric orbit
  (right image), accreting from the clumpy stellar wind of the
  supergiant.  The accretion of matter is persistent in the case of
  the obscured sources, as in the left image, where the compact object
  orbits inside the cocoon of dust enshrouding the whole system. On
  the other hand, the accretion is intermittent in the case of SFXTs,
  which might correspond to a compact object on an eccentric orbit, as
  in the right image.  A 3D animation of these sources is available on the
  website: {\em http://www.aim.univ-paris7.fr/CHATY/Research/hidden.html}
}
\end{figure}

\begin{description}

\item ["Classical" sgHMXBs:]
  the NS is orbiting at a few
  stellar radii only from the star. The absorbed (or obscured) sgHMXBs (like
  IGR\,J16318-4848) are classical sgHMXBs hosting NS constantly
  orbiting inside a cocoon made of dust and/or cold gas, probably
  created by the companion star itself. These systems therefore exhibit a
  persistent X-ray emission.  The cocoon, with an extension of $\sim
  10 \Rstar = 1$\,a.u., is enshrouding the whole binary system. The NS
  has a small and circular orbit (see Figure
  \ref{figure:obscured-sfxt}, left panel).

\item ["Intermediate" SFXT systems:] (such as IGR\,J18483-0311), the
  NS orbits on a small and circular/excentric orbit, and it is only
  when the NS is close enough to the supergiant star that accretion
  takes place, and that X-ray emission arises.

\item ["Classical" SFXTs:] (such as IGR\,J17544-2619), the NS orbits on
  a large and excentric orbit around the supergiant star, and exhibits some
  recurrent and short transient X-ray flares, while it comes close to
  the star, and accretes from clumps of matter coming from the wind of
  the supergiant.  Because it is passing through more diluted medium,
  the $\frac{Lmax}{Lmin}$ ratio is higher for "classical" SFXTs than for
  "intermediate" SFXTs (see Figure \ref{figure:obscured-sfxt}, right
  panel).

\end{description}

Although this scenario seems to describe quite well the characteristics
currently seen in sgHMXBs, we still need to identify the nature of
many more sgHMXBs to confirm it, and in particular the
orbital period and the dependance of the column density with the phase
of the binary system.

\subsection{Formation and evolution of sgHMXBs, link with population synthesis models}

sgHMXBs revealed by {\it INTEGRAL} will allow us to better constrain and understand the formation and evolution of X-ray binary systems, by comparing them to numerical study of LMXB/HMXB population synthesis models. For instance, these new systems might represent a precursor stage of what is known as the "Common Envelope phase" in the evolution of LMXB/HMXB systems, when the orbit has shrunk so much that the neutron star begins to orbit inside the envelope of the supergiant star. In addition, many parameters do influence the various evolutions, from one system to another: differences in mass, size, orbital period, ages, rotation, magnetic field, accretion type, stellar endpoints, etc... Moreover, stellar and circumstellar properties also influence the evolution of high-energy binary systems, made of two massive components likely born in rich star forming regions.

In the very nice review on the formation and evolution of relativistic binaries written by \citet{vandenHeuvel:2009}, the evolution of these supergiant INTEGRAL sources is mentioned. We can have an idea of the formation of these systems, since orbital periods of later evolutionary phases are linearly dependent of initial orbital periods. It is therefore possible to derive that the initial orbital periods of currently very wide O-supergiant INTEGRAL binaries could be as long as 100 days \cite{vandenHeuvel:2009}.
The systems having long orbital periods are expected to survive the
Common Envelope phase. They may then either end as a close eccentric double neutron star, or in some cases, as a black hole-neutron star binary
 (van den Heuvel, priv. comm.).

No such system is known yet, however some of these systems might harbour a black hole as the compact object. Of course neutron stars are easier to detect through X-ray pulsations, but, as Carl Sagan already pointed it out, {\it absence of evidence is not evidence of absence}... We should look for black holes orbiting around supergiant companion stars in wind-accreting HMXBs, however this is only feasible through observational methods involving detection of extremely faint radial velocity displacement due to the high mass of the companion star, or through extremely accurate radio measurements that will be available in the future. On the other hand, massive stars lose so much matter during their evolution that they might always finish as neutron stars (see e.g. \citeauthor{maeder:2008} \citeyear{maeder:2008}). If this is the case, then such systems hosting black holes might not form at all. 

Finally, these sources are also useful to look for massive stellar
"progenitors", for instance giving birth to coalescence of compact
objects, through NS/NS or NS/BH collisions. They would then become
prime candidate for gravitational wave emitters, or even to short/hard
$\gamma$-ray bursts.

\section{Conclusions and perspectives...}

The {\it INTEGRAL} satellite has tripled the total number of Galactic
sgHMXBs, constituted of a NS orbiting around a supergiant
star. Most of these new sources are slow and absorbed X-ray pulsars,
exhibiting a large $\nh$ and long $P_\mathrm{spin}$ ($\sim$1ks).  The
influence of the local absorbing matter on periodic modulations is
different for sgHMXBs or BeHMXBs, segregated in
different parts of $\nh$-$\Porb$ or $\nh$-$P_\mathrm{spin}$.
%
{\it INTEGRAL} revealed 2 new types of sources.  First, the SFXTs, exhibiting short and strong X-ray
flares, with a peak flux of 1 Crab during 1--100s, every $\sim 100$\,days.
These flares can be explained by accretion through
clumpy winds.  Second, the obscured HMXBs are persistent X-ray sources
composed of supergiant stellar companions exhibiting a strong intrinsic absorption and long $P_\mathrm{spin}$.  The NS is
deeply embedded in the dense stellar wind, forming a dust cocoon
enshrouding the whole binary system.

These results show the existence in our Galaxy of a dominant
population of a previously rare class of high-energy binary systems:
sgHMXBs, some of them exhibiting a high intrinsic absorption
(\citeauthor{chaty:2008} \citeyear{chaty:2008};
\citeauthor{rahoui:2008} \citeyear{rahoui:2008}). Studying
this population will provide a
better understanding of the formation and evolution of short-living
HMXBs.  Furthermore, stellar population models now have to
take these objects into account, to assess a realistic number of
high-energy binary systems in our Galaxy. 




\begin{theacknowledgments}
  First, I would like to thank the organisers, and especially Vicky Kalogera,
  for such a successfully organized and nice workshop, in a nice place,
  ideal for new ideas to appear! 
Then, I thank here Ed van den Heuvel and Philip Podsialowski for nice and useful discussions on INTEGRAL sources.
  Finally, I am endlessly grateful to all my close collaborators: 
  A. Bodaghee, Q.Z. Liu, I. Negueruela, L. Pellizza, 
  F. Rahoui, J. Rodriguez, J. Tomsick, J.Z. Yan, J. A. Zurita Heras, 
  and also P. Filliatre, P.-O. Lagage and R. Walter 
  for many fruitful work and discussions on the study of {\it INTEGRAL} sources.
This work was supported by the Centre National d'Etudes Spatiales (CNES), based on observations obtained with MINE --the Multi-wavelength INTEGRAL NEtwork--.
\end{theacknowledgments}



\bibliographystyle{aipproc}   


\input{chaty_mykonos.bbl}
\IfFileExists{\jobname.bbl}{}
 {\typeout{}
  \typeout{******************************************}
  \typeout{** Please run "bibtex \jobname" to optain}
  \typeout{** the bibliography and then re-run LaTeX}
  \typeout{** twice to fix the references!}
  \typeout{******************************************}
  \typeout{}
 }

\end{document}

%% file: symbols.tex
\def\cmmoinsdeux{\mbox{ cm}^{-2}}

\def\microns{\mbox{ } \mu \mbox{m}}

\def\kms{\mbox{ km\,s}^{-1}}

\def\Mdot{\frac{dM}{dt}}
\def\Msol{\mbox{ }M_{\odot}}
\def\Rsol{\mbox{ }R_{\odot}}
\def\Lsol{\mbox{ }L_{\odot}}

\def\Rstar{\mbox{ }R_{\star}}
\def\Tstar{\mbox{ }T_{\star}}

\def\Porb{P_{orb}}

\def\mags{\mbox{ magnitudes}}

\def\ergs{\mbox{ erg\,s}^{-1}}

\def\adeg{^{\circ}}

\def\amin{^\prime}

\def\nh{N_{\rm H}}

\def\ltsima{\; \buildrel < \over \sim \;}
\def\simlt{\lower.5ex\hbox{\ltsima}}            
\def\gtsima{\; \buildrel > \over \sim \;}
\def\simgt{\lower.5ex\hbox{\gtsima}}            
